\documentclass[aps,prd,superscriptaddress,11pt,showpacs,notitlepage]{revtex4}
	\usepackage{graphicx}
	\usepackage{amsmath,amssymb,mathrsfs}
\usepackage{epsf}
\usepackage{epsfig}

\usepackage{amsmath}

\usepackage{amssymb}

\numberwithin{equation}{section}

\newcommand{\be}{\begin{equation}}
\newcommand{\ee}{\end{equation}}
\newcommand{\bea}{\begin{eqnarray}}
\newcommand{\eea}{\end{eqnarray}}

\newcommand{\bb}{\bibitem}
 
\newcommand{\eqn}{\begin{eqnarray}}
\newcommand{\eqnx}{\end{eqnarray}}

\begin{document}
\title{Topological phase transitions in the gauged BPS baby Skyrme model}
\author{C. Adam}
\affiliation{Departamento de F\'isica de Part\'iculas, Universidad de Santiago de Compostela and Instituto Galego de F\'isica de Altas Enerxias (IGFAE) E-15782 Santiago de Compostela, Spain}
\author{C. Naya}
\affiliation{Departamento de F\'isica de Part\'iculas, Universidad de Santiago de Compostela and Instituto Galego de F\'isica de Altas Enerxias (IGFAE) E-15782 Santiago de Compostela, Spain}
\author{T. Romanczukiewicz}
\affiliation{Institute of Physics,  Jagiellonian University,
Lojasiecza 11, 30-348 Krak\'{o}w, Poland}
\author{J. Sanchez-Guillen}
\affiliation{Departamento de F\'isica de Part\'iculas, Universidad de Santiago de Compostela and Instituto Galego de F\'isica de Altas Enerxias (IGFAE) E-15782 Santiago de Compostela, Spain}
\author{A. Wereszczynski}
\affiliation{Institute of Physics,  Jagiellonian University,
Lojasiecza 11, 30-348 Krak\'{o}w, Poland}

\begin{abstract}
We demonstrate that the gauged BPS baby Skyrme model with a double vacuum potential allows for phase transitions from a non-solitonic to a solitonic phase, where the latter corresponds to a ferromagnetic liquid. Such a transition can be generated by increasing the external pressure $P$ or by turning on an external magnetic field $H$. As a consequence, the topological phase where gauged BPS baby skyrmions exist, is a higher density phase. For smaller densities, obtained for smaller values of $P$ and $H$, a phase without solitons is reached. We find the critical line in the $P,H$ parameter space. Furthermore, in the soliton phase, we find the equation of state for the baby skyrmion matter $V=V(P,H)$ at zero temperature, where $V$ is the "volume", i.e., area of the solitons.
\end{abstract}
\maketitle 
\section{Introduction}
The baby Skyrme model \cite{baby} (see also \cite{karliner}-\cite{shnir}) is a lower dimensional version of the $3+1$ dimensional Skyrme model \cite{skyrme}, that is, an effective theory of the low energy, nonperturbative sector of Quantum Chromodynamics. It plays an important role as a toy model for the higher dimensional theory. In fact, many aspects of the original Skyrme proposal were understood qualitatively by the investigation of its $2+1$ dimensional counterpart. Recently, it has been understood that the correct Skyrme model should be a near BPS theory \cite{nearBPS}, \cite{SutBPS} where the BPS part  gives the main contribution to some static observables \cite{BPS} (see also \cite{Marl}, \cite{Sp2}). A concrete near BPS Skyrme action has been proposed in \cite{nearBPS}. Indeed, already its BPS part leads to very accurate results not only for the binding energies of the atomic nuclei \cite{nearBPS} and some thermodynamical properties \cite{term} of the nuclear matter but also for the physics of the neutron stars \cite{star} (maximal masses and radii, equation of state). There are, therefore, strong arguments for the assumption that this near BPS Skyrme model, providing an unified description of the nuclear matter from baryons with the baryon charge $B=1$ and nuclei $B\sim 10^2$ to neutron stars $B \sim 10^{57}$, is a good candidate for the correct low-energy effective model of QCD. 
\\
Qualitatively, the BPS Skyrme model consists of two terms: the topological (baryon) current squared and a potential.  Obviously, there is a baby version of the BPS Skyrme model \cite{GP}, \cite{restr-bS}, \cite{Sp1}, \cite{stepien} (for some near BPS extension see \cite{near-baby}). Both theories have many features in common: 1) the existence of the linear energy - topological charge bound which is saturated by infinitely many solutions with arbitrary topological charge; 2) the invariance of the static energy integral under the SDiff symmetries and the resulting close relation to an incompressible liquid as the corresponding state of matter; 3) the existence of a standard thermodynamics with equivalence between thermodynamical and field theoretical pressure.  
\\
Recently, the baby BPS Skyrme model has been used to investigate magnetic properties of skyrmions after minimal coupling to the Maxwell electrodynamics \cite{BPS-g}, \cite{BPS-g2}. Interestingly, the gauged baby BPS Skyrme model also allows for a $N=2$ supersymmetric extension, see, e.g., \cite{susyBPS}, \cite{nitta-susy}, \cite{near-baby}. Undoubtedly, the interaction with an abelian gauge field is very important for properties of baryons and atomic nuclei. It is a well-known result by Witten that, although the precise form of the low energy effective Skyrme-type theory is still unknown, its coupling to the electromagnetic $U(1)$ field is completely fixed \cite{wit1}. Hence, in principle, one can study electromagnetic properties of skyrmions once a particular model is chosen. In practice, however, this is a rather complicated task even within the framework of the BPS Skyrme model, and not many results are available (for the standard Skyrme model see \cite{g-Skyrme1}-\cite{g-Skyrme2} and \cite{foster}).  Again, the BPS baby Skyrme model serves here as a perfect laboratory (for gauged solitons in a related Skyrme-Faddeev-Niemi model see \cite{shnir1}). 
\\
The aim of the present paper is to further investigate the gauged BPS baby Skyrme model \cite{BPS-g}, \cite{BPS-g2}. Specifically, we would like to analyze the issue of the (non)existence of planar skyrmions in the model with double vacuum potentials. It is one of the most surprising findings in our former work that, in contrast to one vacuum potentials, the gauged BPS baby Skyrme model does not support solitons for potentials with two vacua. This unexpected observation (there is no such difference for the non-gauged case, see for example recent works for the Skyrme model \cite{bjarke}) was made for BPS configurations, i.e., with zero external pressure and no external magnetic field. Here we will show that solitons can exist if these two {\it external} parameters are larger than some minimal {\it critical} values. Hence a phase transition between a solitonic phase and a non-solitonic phase is observed.  
\section{The gauged BPS baby Skyrme model}
The gauged BPS baby Skyrme model is given by the following Lagrangian density
\begin{equation}\label{Lag}
 {\cal L} = -\frac{\lambda^2}{4} \left( D_\mu \vec \phi \times D_\nu  \vec \phi \right)^2 - \mu^2 U ( \vec n \cdot \vec \phi)
 + \frac{1}{4 g^2} F^2_{\mu \nu}
\end{equation}
where the covariant derivative reads  \cite{GPS}, \cite{schr1}
\be
D_\mu \vec{\phi}\equiv \partial_\mu \vec{\phi} +A_\mu \vec{n} \times \vec{\phi} .
\ee
As usual, without loss of generality, we assume that the constant vector $\vec{n}=(0,0,1)$ and the potential $U$ is a function of the third component of the unit vector field $\vec{\phi}=(\phi_1,\phi_2,\phi_3)$. The gauge field is the Maxwell field with the electromagnetic coupling constant $g$. Here we list the main properties of this model.
\begin{itemize}
\item[] {\it i) BPS bound} 
\\
The static energy (or regularized energy for a non-zero constant external magnetic field $H$) 
\be
E_{reg}= \frac{E_0}{2} \int d^2 x \left[ \lambda^2 \left( D_1\vec{\phi} \times D_2 \vec{\phi}\right)^2 +2\mu^2 U +\frac{1}{g^2} (B-H)^2
\right]
\ee
is bounded from below by the topological charge $k$
\begin{equation}
E_{reg} \geq 4\pi |k| E_0 \lambda^2 <W'>_{S^2} .
\end{equation}
Here $B$ is the magnetic field (pseudoscalar in 2+1 dimensions), $W$ is the so-called superpotential (depending on the target space variable $\phi_3$) defined as
 \begin{equation} \label{superpot-eq}
\lambda^2 W'^2 + g^2\lambda^4 W^2 -2\lambda^2WH =2\mu^2 U ,
\end{equation}
and $ <W'>_{S^2} $ is its average value over the target space manifold. The prime denotes differentiation with respect to $\phi_3$.
\item[] {\it ii) BPS equations}
\\
The bound is saturated by solutions of the BPS equations
\begin{equation}
Q= W' \label{bps1}
\end{equation}
\begin{equation}
B = - g^2\lambda^2W +H \label{bps2}
\end{equation}
where 
\begin{equation}
Q = q+\epsilon_{ij} A_i \partial_j (\vec{n} \cdot \vec{\phi} ), \;\;\; q = \vec{\phi} \cdot \partial_1 \vec{\phi} \times \partial_2 \vec{\phi} .
\end{equation}
and $q$ is the topological charge density. 
\item[] {\it iii) No solitons for two-vacuum potentials}
\\
The existence of such solutions is a rather subtle issue as the superpotential $W$ has to be defined over the whole interval $\phi_3 \in [-1,1]$. This requirement is very restrictive and, for example, {\it cannot be satisfied} in the case of two vacuum potentials and $H=0$. The fact that the BPS baby skyrmions (two vacuum potential) can be destabilized by the electromagnetic interaction was quite surprising since no similar effect was known in the non-gauged  case. 
\item[] {\it iv) Magnetization} 
\\
The thermodynamic magnetization M, defined as minus the change of the thermodynamic energy of a sample (in our case, the skyrmion) under a variation of the external magnetic field
\be
M= - \frac{\partial E_{reg}}{\partial H}
\ee
exactly agrees with the usual definition of magnetization as the
difference between full and external magnetic flux in the sample
\be
M=\frac{1}{g^2} \int (B-H) d^2x .
\ee
Furthermore, the matter described by the gauged BPS Skyrme model behaves as a rather {\it nonlinear ferromagnetic medium} where the solitons remain magnetized even when the external magnetic field vanishes, i.e., they possess a permanent magnetization. The external magnetic field $H$ acts in two ways: it change the size of the baby skyrmions (it squeezes solitons if has the same sign like the permanent magnetization M of the skyrmion, while it enlarges the skyrmion if $H$ and $M$ have opposite signs) and interferes with the magnetic field inside the solitons: for an external magnetic field $H$ which is sufficiently weak and oppositely oriented to $M$, the phenomenon of magnetic flux inversion occurs. That is, the total magnetic field $B$ flips sign in a shell region near the boundary of the skyrmion, as it has to take the value $B = H$ at the boundary. On the other hand, it preserves its original sign resulting from the permanent magnetization in the interior (core region) of the skyrmion.
\item[] {\it v) Non-BPS (external pressure) solutions}
\\
Solutions with a non-zero {\it field theoretical pressure} ($P=-T_{ii}$, where $T_{\mu \nu}$ is the energy-momentum tensor with the constant external magnetic part subtracted) are described by the first order BPS equations but with a modified superpotential equation
\be
\lambda^2W'^2+g^2\lambda^4W^2 - 2\lambda^2 WH=2\mu^2U +2P.
\ee
The pressure introduced in this way obeys the usual thermodynamical relation
\be
P=-\frac{\partial E_{reg}}{\partial V}
\ee
where $V$ is the geometrical (compacton) "volume" (area) and therefore agrees with the thermodynamical pressure. 
\end{itemize}
\section{The new baby potential}
In order to understand how the external magnetic field and/or pressure may influence the existence of planar solitons in the case of two vacuum potentials we have to analyze 
the necessary condition for the existence of BPS gauged baby skyrmions, that is, the existence of a global solution to the superpotential equation
\be
 \lambda^2 W'^2+g^2\lambda^4 W^2-2\lambda^2 W H =2\mu^2 U +2P.
\ee
Concretely, we shall consider the so-called new baby potential
\be
U=\frac{1}{4}(1-\phi_3^2) .
\ee
The next step (if the superpotential exists) is to solve the corresponding equations of motion. Here we assume the standard axially symmetric static ansatz
\begin{equation} \label{rad-ans}
\vec{\phi} (r,\phi)  = \left( 
\begin{array}{c}
\sin f(r) \cos n\phi \\
\sin f(r) \sin n\phi \\
\cos f(r)
\end{array}
\right), \;\;\;\; A_0=A_r=0, \;\;\; A_\phi=na(r) 
\end{equation}
which leads to an identically vanishing electric field and to the magnetic field $B=\frac{na'(r)}{r}$. Moreover, we introduce a new base space variable $y=r^2/2$ and target space variable $h$ via
\begin{equation}
\phi_3 = \cos f \equiv 1-2h \; \Rightarrow \; h =\frac{1}{2} (1-\cos f), \;\;\; h_y = \frac{1}{2} \sin f f_y.
\end{equation} 
Then the first order field equations (BPS equations) read
\begin{equation}
2nh_y(1+a)=-\frac{1}{2}W_h
\end{equation}
\begin{equation}
na_y=-g^2\lambda^2 W +H
\end{equation}
where the following boundary condition 
\be
h(0)=1, \;\;\; h(y_P)=0,
\ee
\be
a(0)=0, \;\;\; a_y(y_P)=\frac{H}{n}
\ee
guarantee the nontrivial topology ($n$ is just the topological charge) and finiteness of the (regularized) energy
\be
 E_{reg} =2\pi \int dy \left(2\lambda^2n^2(1+a)^2h_y^2+\mu^2 h(1-h)+\frac{1}{2g^2}n^2a_y^2 \right).
\ee 
Here, $y_P$ is the size of the compacton for a non-zero value of the pressure $P$. Since we deal with compact solutions one can define the geometrical volume (area) 
\be
V=2\pi y_P 
\ee
which, as we remarked before, together with the pressure fulfills the usual thermodynamical relation. 
The superpotential equation in terms of the new variable $h$ reads
\be
 \frac{\lambda^2}{4} W_h^2+g^2\lambda^4 W^2-2\lambda^2 W H =2\mu^2 U +2P.
\ee
Further, we always assume that the Skyrme field approaches the vacuum at $h=0$ at the compacton radius $r_P = \sqrt{2y_P}$. This, together with the boundary conditions and BPS equations for $h$ and $a$, implies that the superpotential $W(h)$ must obey the boundary conditions
\be
W(0)=0, \;\;\; W_h (0)=\pm \frac{2}{\lambda} \sqrt{2P}.
\ee
For nonzero pressure this implies that the first derivative $h_y$ jumps at the compacton boundary. This is as it should be, because if we impose a nonzero external pressure at the soliton surface (compacton boundary), then the energy density must jump there. 

In numerical computations we assumed $\lambda=1$ and $\mu^2=0.1$, while the electromagnetic coupling constant was kept as a free parameter. Furthermore, a particular value of the topological charge $n$ can be eliminated from the field equation by the coordinate redefinition $\tilde{y} = y/n$ rendering the numerics $n$ independent (with a linear dependence of the energy and volume on $n$ - as expected for a BPS model). Therefore we present all plots for $n=1$. 
\subsection{$H=0, P=0$}
\begin{figure}
\includegraphics[height=5.5cm]{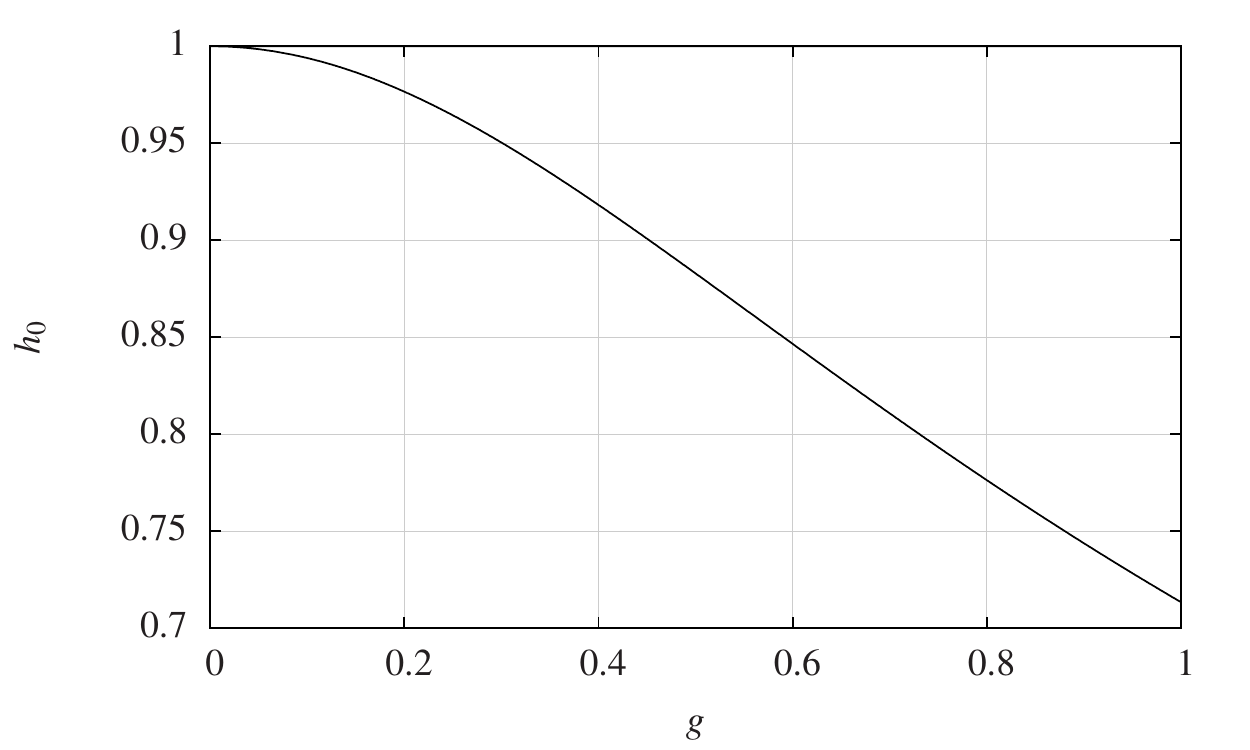}
\label{sing}
\caption{The position of the singularity in the superpotential $W$ for $H=0$ and $P=0$, as a function of the coupling constant $g$.}
\end{figure}

Let us begin with the case where $H=0$, $P=0$. It is straightforward to see that we arrive at some difficulties. Indeed, the superpotential equation 
 \be
\frac{\lambda^2}{4} W_h^2+g^2\lambda^4 W^2=2\mu^2 h(1-h) \label{susyN}
\ee
enforces boundary conditions at both ends of the unit interval.  In addition to $W(0)=0$, we must impose $W(1)=0$. However, for a first order equation it is in general impossible that both conditions can be simultaneously obeyed. Indeed, if we numerical solve equation (\ref{susyN}) then a singularity occurs for $h=h_0$. (Strictly speaking $W_h(h_0)=0$ while $W_{hh} (h_0)=\infty$.) It approaches the boundary of the interval as the electromagnetic coupling constant goes to 0, see Fig. 1. Then, for $g=0$ we arrive at the non-gauge case for which BPS skyrmions do exist and read
\be
h(y) = \left\{
\begin{array}{lc}
\sin^2 \frac{\pi}{2} \left( 1 - \frac{y}{y_0} \right) & y\leq y_0 \\
0 & y \geq y_0
\end{array} \right., \;\;\;\;\;\;\; \mbox{where} \;\;\;\;\; y_0=\frac{n\pi \lambda}{4\mu} .
\ee
Obviously, the nonexistence of the superpotential shows that gauged solitons do not show up if the pressure as well as the external magnetic field vanish (for any value of $g>0$). It is also clearly visible that in the more general situation, with nonzero value for the external parameters $P$ and $H$, the superpotential might exist on the full unit interval for some external parameter values. 
\subsection{$H=0, P>0$}
\begin{figure}
\includegraphics[height=6.5cm]{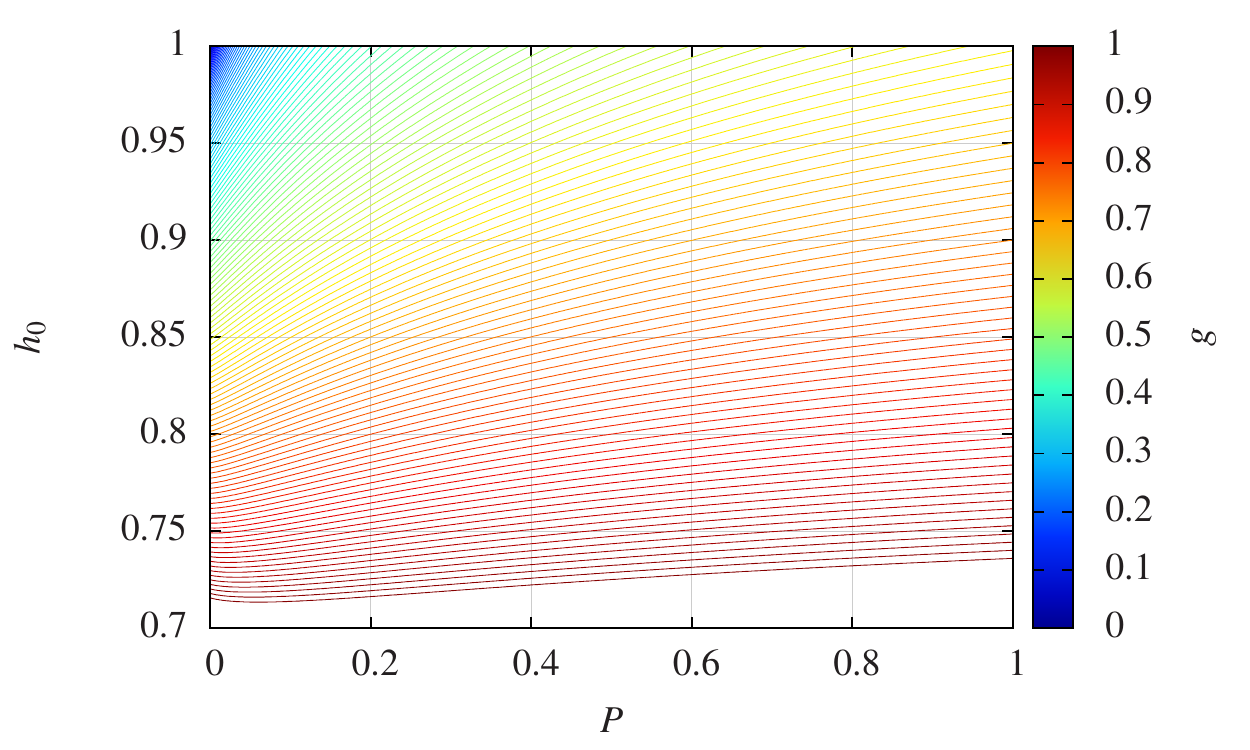}
\caption{The position of the singularity for the superpotential for $H=0$. For $h_0\ge 1$ solutions do exist.}
\end{figure}
Now let us turn on the pressure, still assuming $H=0$, at the boundary of compactons. For the limiting case, when $g=0$, the exact solutions are known \cite{BPS-g2} 
\be
\frac{1-2h}{2\sqrt{\frac{P}{\mu^2} +h(1-h)}} = \tan \frac{\mu}{n\sqrt{2} \lambda} (y - y_0), \;\;\; \mbox{where} \;\;\; \tan  \frac{\mu y_0}{n\sqrt{2} \lambda} = \frac{\mu}{\sqrt{P}}
\ee
leading to the following equation of state
\be
V=\frac{4\sqrt{2} \pi \lambda n}{\mu} \arctan \frac{\mu}{\sqrt{P}} .
\ee
Hence, for $P\rightarrow 0$ we find the maximal volume $V_{max}=\frac{2\sqrt{2} \pi^2 \lambda n}{\mu}$. For higher charge solutions the volume can be further increased by taking a collection of smaller charge constituents with an empty space (vacuum) in between. Such a phase resembles a gas of BPS skyrmions. On the other hand, for $V < V_{max}$ we have a sort of liquid phase. Therefore, one can conclude that we observe a gas-liquid phase transition at $P=0$ (and no electromagnetic coupling). In fact, such a behavior is a generic feature of the BPS Skyrme models in any dimensions \cite{term}. 
\\  
If the electromagnetic coupling constant is not zero then solutions exist only for $P \ge P_{min}$. At the critical minimal pressure $P_{min}$ the system reaches its maximal volume $V_{max}(P)$. For $P < P_{min}$ the superpotential develops a singularity for $h_0 <1$, see Fig. 2, and solutions cease to exist. Hence, at $P_{min}$ another phase transition occurs between a topologically trivial phase (no solitons) and a skyrmionic matter phase. One can also say that the gauged BPS skyrmions are pressure induced objects. In Fig. 3, we plot the equation of state $V=V(P)$ for the skyrmionic phase for several values of the electromagnetic coupling constant. The ends of the curves correspond to $P_{min}$ and $V_{min}$. A detailed plot is presented in Fig. 4, where the line of minimal pressure is visible as a function of $g$. 
\begin{figure}
\includegraphics[height=6.5cm]{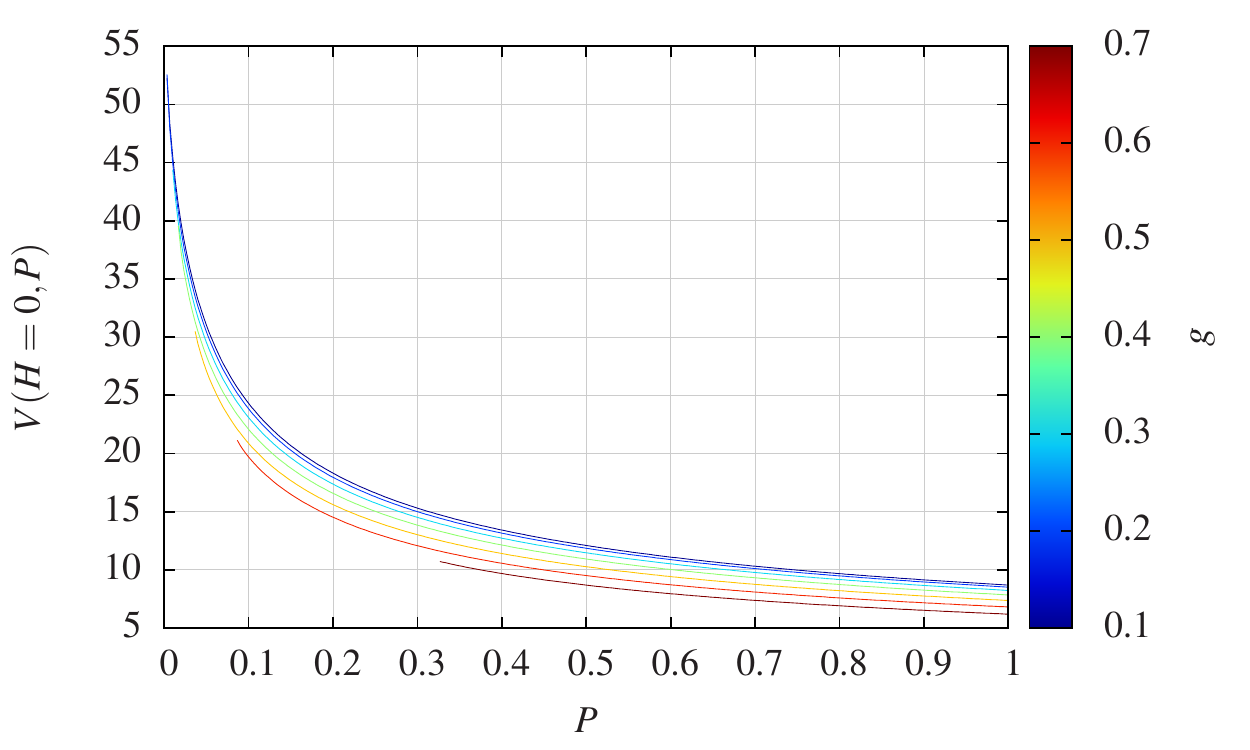}
\caption{Volume vs. pressure, for different values of the coupling constant $g$.}
\end{figure}
\begin{figure}
\includegraphics[height=6.5cm]{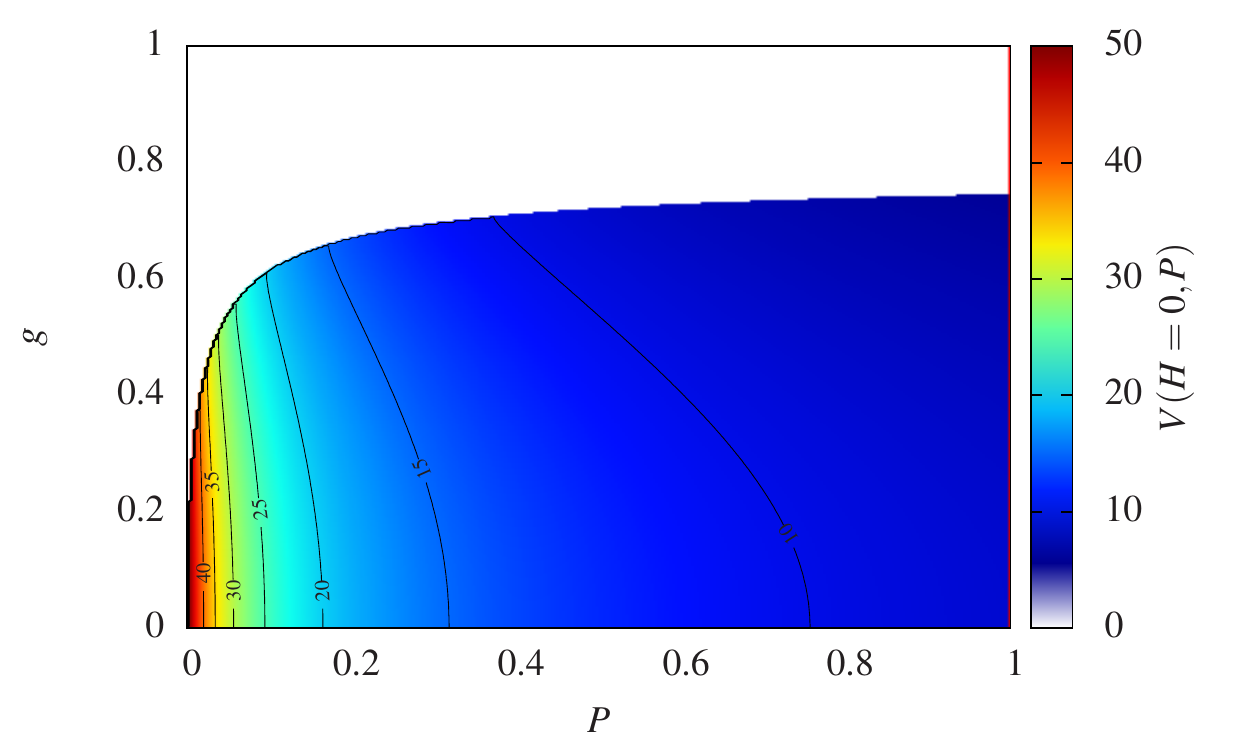}
\caption{The volume as a function of pressure $P$ and coupling constant $g$. The white region denotes no solutions.}
\end{figure}
\subsection{$H \neq 0, P=0$}
\begin{figure}
\includegraphics[height=6.5cm]{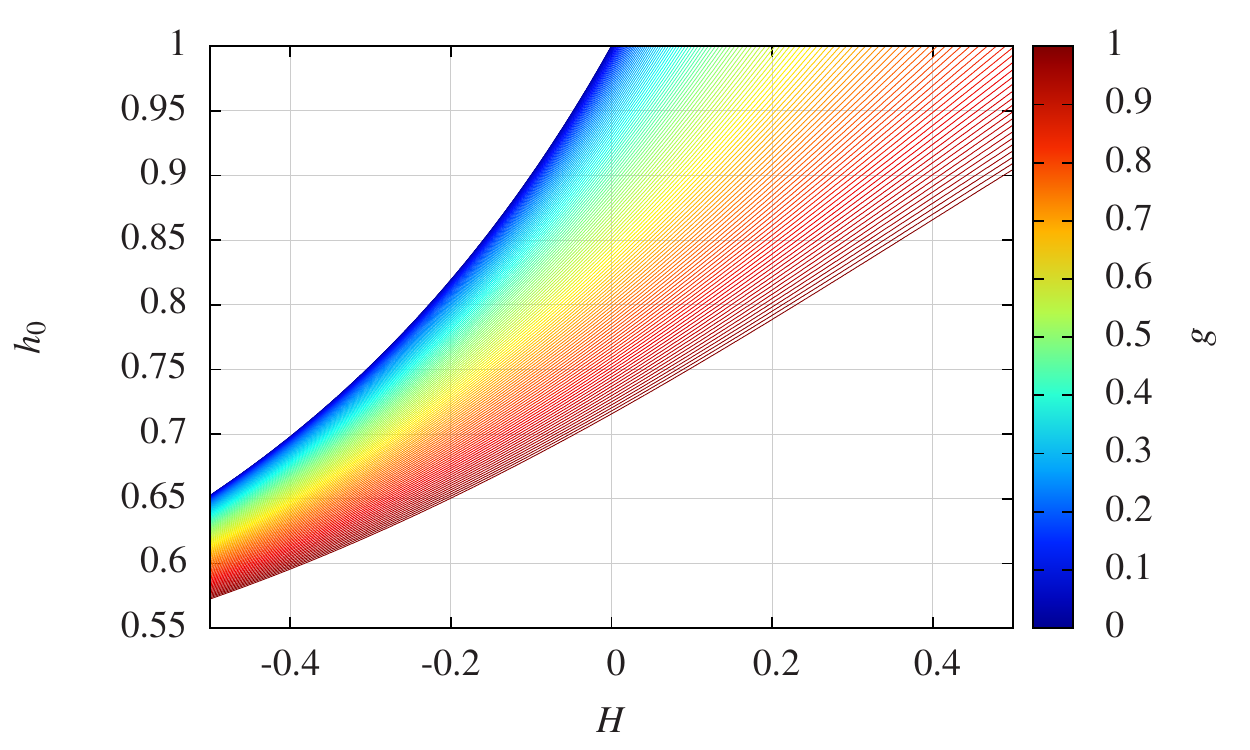}
\caption{The position of the singularity for the superpotential for $P=0$. For $h_0\ge 1$ solutions do exist.}
\end{figure}
\begin{figure}
\includegraphics[height=6.5cm]{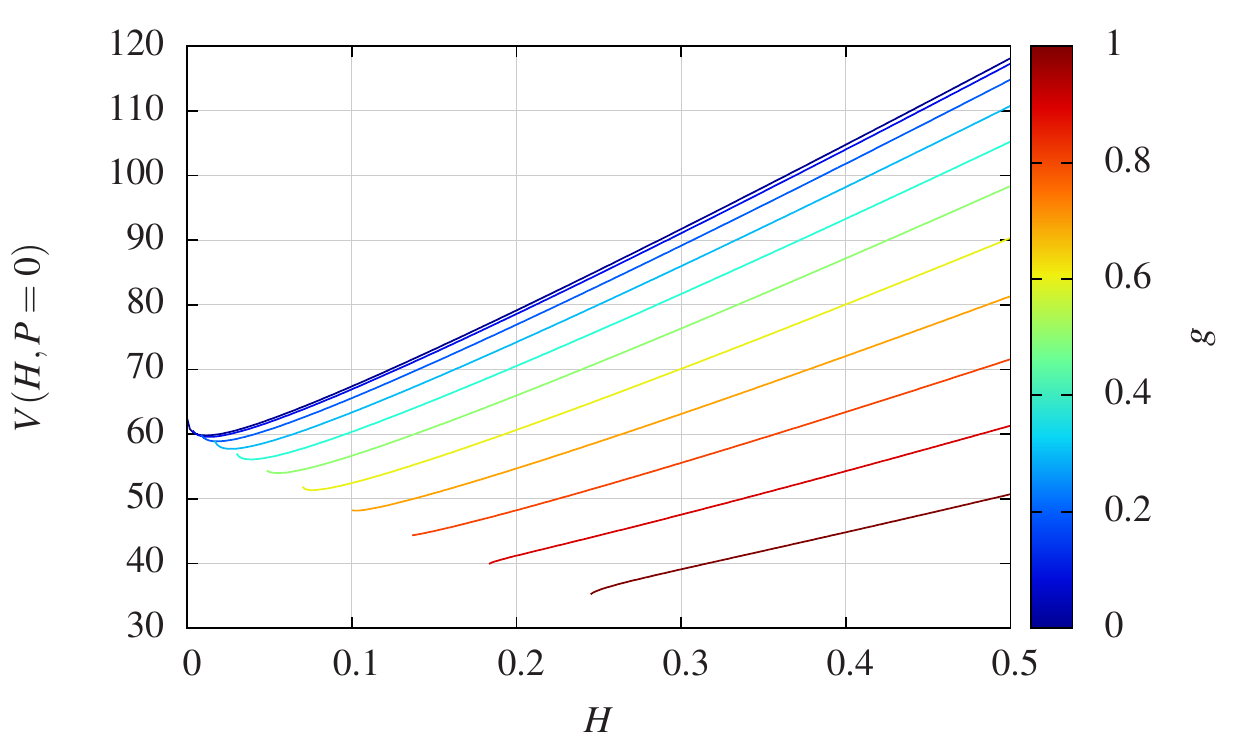}
\caption{Volume vs. pressure, for different values of the coupling constant $g$.}
\end{figure}
\begin{figure}
\includegraphics[height=6.5cm]{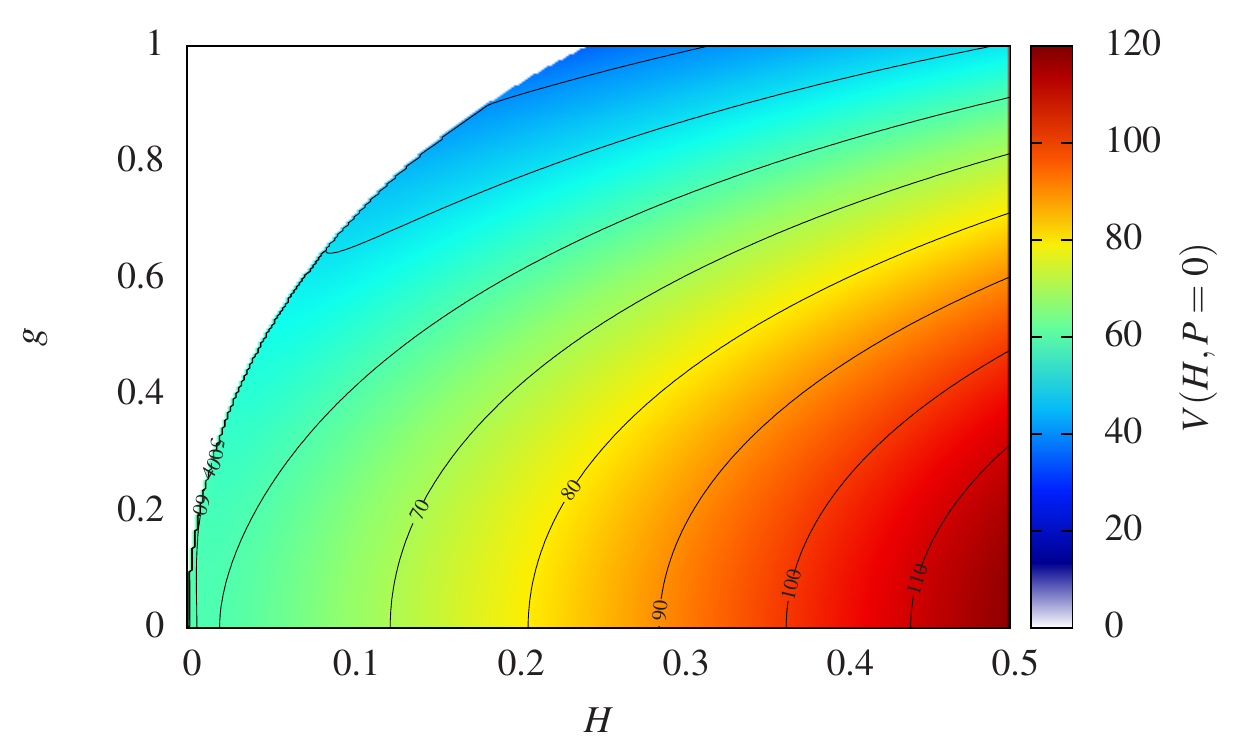}
\caption{The volume as a function of the external magnetic field $H$ and coupling constant $g$. The white region denotes no solutions.}
\end{figure}
Another possibility to create solitons, i.e., to extend the existence of the superpotential on the whole unit interval, is to consider a non-vanishing external magnetic field $H>0$ together with $P=0$,
\be
\frac{\lambda^2}{4} W_h^2 +g^2 \lambda^2 W^2=2\mu^2 U +2\lambda^2 W H .
\ee
Let us start with the limiting case when $g=0$. The corresponding superpotential equation is 
\be
\frac{\lambda^2}{4} W_h^2 =2\mu^2 U +2\lambda^2 W H 
\ee
It is not difficult to show that $W_h$ will never take the value zero on the unit segment (except at $h=0$). Indeed, assuming the positive square root we find that $W_h >0$ in the vicinity of $h=0$. Then, the right hand side of the equation is always greater than 0 as $U \ge 0$ and $W$ is a growing function. A consequence of that is the existence of solitons for any $H$ (for $g=0$ and $P=0$). Again, if $g>0$, the original ($H=0$) singularity moves towards the boundary of the unit segment, reaching it for a certain minimal value of the magnetic field $H_{min}$, Fig. 5. Starting with this minimal value, the external magnetic field creates a skyrmionic phase. Now, the volume increases with the magnetic field, Fig. 6, as the magnetization inside the solitons is opposite to the orientation of the external filed. A more detailed equation of state $V=V(H)$ for different $g$ is presented in Fig. 7.
\\
It is a surprising feature of the gauged BPS baby Skyrme model that the limiting case, $g=0$, can be solved exactly although the superpotential is not known analytically. Let us write the full second order field equation for the Skyrme field assuming the non-back reaction approximation (equivalent to $g=0$), i.e., $a=\beta y$ where $\beta \equiv H/n$.  Then,
\be
\partial_y \left[ h_y (1+a)^2 \right] = \frac{\mu^2}{4n^2\lambda^2} (1-2h)
\;\; \Rightarrow \;\; \partial_y \left[ h_y (1+\beta y)^2 \right] = \frac{\mu^2}{4n^2\lambda^2} (1-2h)
\ee
Now we introduce the new variable
\be
z = \frac{1}{1+\beta y} .
\ee
Then,
\be
z^2 h_{zz}=\frac{\mu^2}{4n^2\lambda^2 \beta^2} (1-2h) \equiv \alpha (1-2h)
\ee
with the following general solution: for $\alpha < \frac{1}{8}$
\be
h(z)=C_1z^{\frac{\sqrt{1-8\alpha}+1}{2}}+C_2z^{\frac{-\sqrt{1-8 \alpha}+1}{2}} +\frac{1}{2},
\ee
for $\alpha > \frac{1}{8}$
\be
h(z) = C_1  \sqrt{z} \sin \left( \frac{\sqrt{8 \alpha-1}}{2} \ln z \right) +  C_2  \sqrt{z} \cos \left( \frac{\sqrt{8 \alpha-1}}{2} \ln z \right)   +\frac{1}{2} .
\ee
The boundary conditions
\be
h(z=1)=1, \;\; h(z_0)=0, \;\; h_z(z_0)=0
\ee
lead to the following solution ($\alpha <1/8$)
\be
h(z)=\frac{1}{2} -\frac{\sqrt{z}}{2} \left( \frac{1}{\sqrt{z_0} \sin (\omega \ln z_0 )} +\cot (\omega \ln z_0) \right) \sin (\omega \ln z) +\frac{1}{2} \sqrt{z} \cos (\omega \ln z),
\ee
where $\omega \equiv \sqrt{8\alpha -1}$ and the position of the compacton boundary is given by
\be
\sin (\omega \ln z_0) + \omega \cos (\omega \ln z_0) = -2\omega \sqrt{z_0} .
\ee
Hence, we find the equation of state relating the volume $V=2\pi y_0$ and the magnetic field
\be
 \sin \left[ \frac{1}{2} \sqrt{\frac{\mu^2}{2H^2\lambda^2}-1}   \ln \left( 1+\frac{HV}{2\pi n} \right)  \right] 
+2  \sqrt{\frac{\mu^2}{2H^2\lambda^2}-1}\cos \left[ \frac{1}{2} \sqrt{\frac{\mu^2}{2H^2\lambda^2}-1}   \ln \left( 1+\frac{HV}{2\pi n} \right)   \right]   
\nonumber
\ee
\be
\hspace*{9cm}  =  -2 \sqrt{ \frac{ \frac{\mu^2}{2H^2\lambda^2}-1}{ 1+\frac{HV}{2\pi n }}}  
\ee
for the external magnetic field obeying 
\be
H < \frac{\mu}{\lambda\sqrt{2}} .
\ee
If the magnetic field is bigger than this value one has to change to the hyperbolic functions resulting in
\be
 -\sinh \left[ \frac{1}{2} \sqrt{1-\frac{\mu^2}{2H^2\lambda^2}}   \ln \left( 1+\frac{HV}{2\pi n} \right)  \right]
+2  \sqrt{1-\frac{\mu^2}{2H^2\lambda^2}}\cosh \left[ \frac{1}{2} \sqrt{1-\frac{\mu^2}{2H^2\lambda^2}}   \ln \left( 1+\frac{HV}{2\pi n} \right)   \right]   \nonumber
\ee
\be
\hspace*{9.0cm} =  -2 \sqrt{ \frac{ 1-\frac{\mu^2}{2H^2\lambda^2}}{ 1+\frac{HV}{2\pi n }}} .
\ee
These results give us the approximated equation of state $V=V(H)$ for small value of the electromagnetic coupling constant. 
\subsection{$H \neq 0, P \neq 0$}
\begin{figure}
\includegraphics[height=6.5cm]{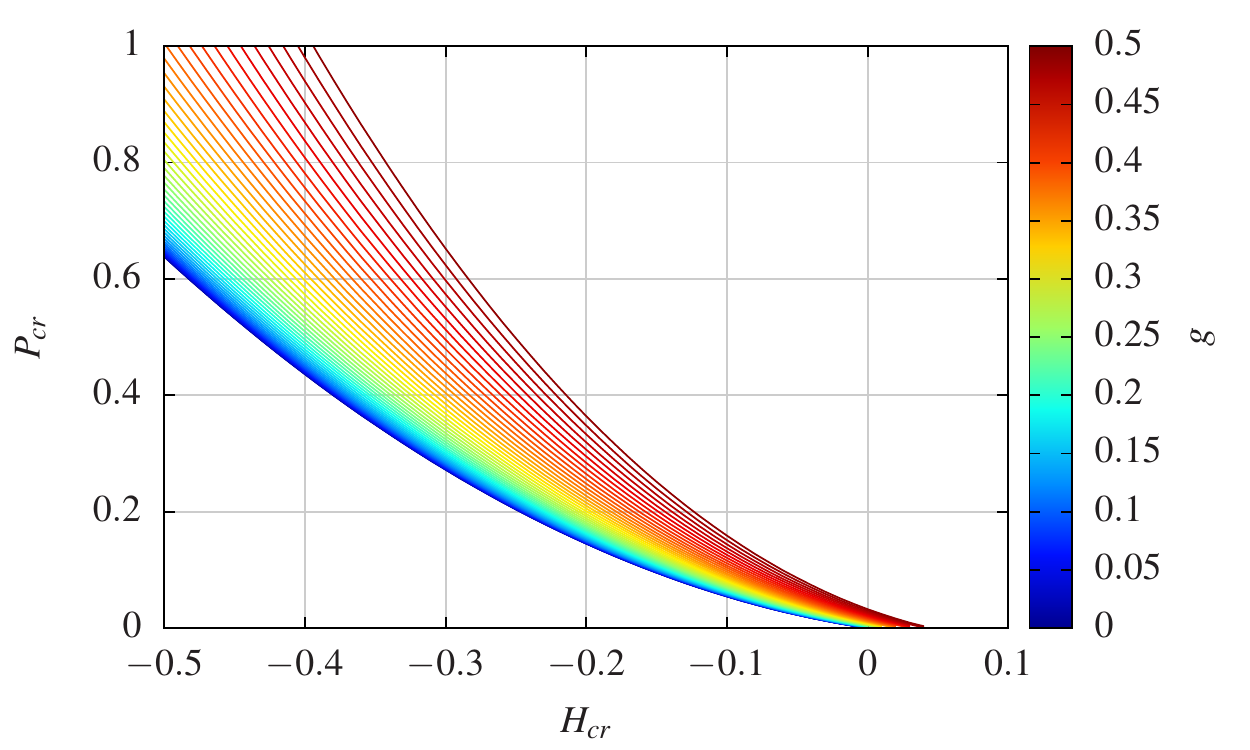}
\caption{Critical line for different values of the coupling constant $g$. Above the line the topological phase exists while below the line we have the non-topological phase.}
\end{figure}
\begin{figure}
\includegraphics[height=6.5cm]{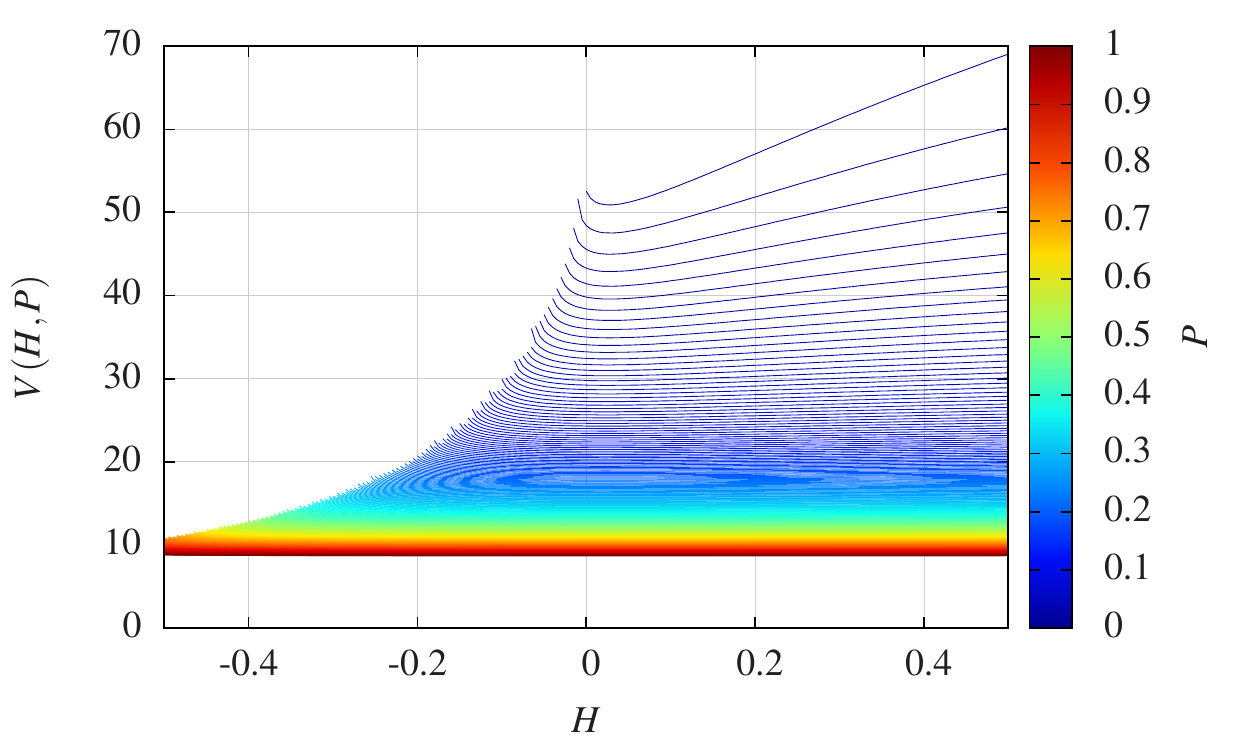}
\caption{The volume as a function of $P$ and  $H$. Here $g=0.2$}
\end{figure}
In the most general situation, one imposes the external pressure as well as the external magnetic field. Again, solitons are observed for sufficiently large $P$ and $H$. In fact, one can plot a critical curve in the $P, H$ plane above which the solitonic phase exists, see Fig. 8. The corresponding equation of state $V=V(H,P)$, for $g=0.2$, is presented in Fig. 9. The dependence of the regularized energy, as well as its derivative, on the magnetic field $H$ for several values of $P$ (and $g=0.2$) are presented in Fig. 10.
\\
The soliton medium possesses a permanent magnetization, that is, a non-trivial magnetic field inside the solitons, even for vanishing external magnetic field. Of course, this makes sense only if such solitons do survive in the $H=0$ limit, which  requires a sufficiently large non-vanishing external pressure. We plot the magnetization in Fig. 11. It is always negative and grows to a constant value in the limit of very large pressure. In Fig. 12 we show the magnetic susceptibility at $H=0$. There is a qualitative different to the one-vacuum potential case: although the regularized energy is a monotonously growing function with $H$ (positive first derivative and negative magnetization), it is a concave function (with negative second derivative respect to $H$). This happens for any $P$. Hence, the magnetic susceptibility (at $H=0$) is always positive,
\begin{figure}
\includegraphics[height=4.5cm]{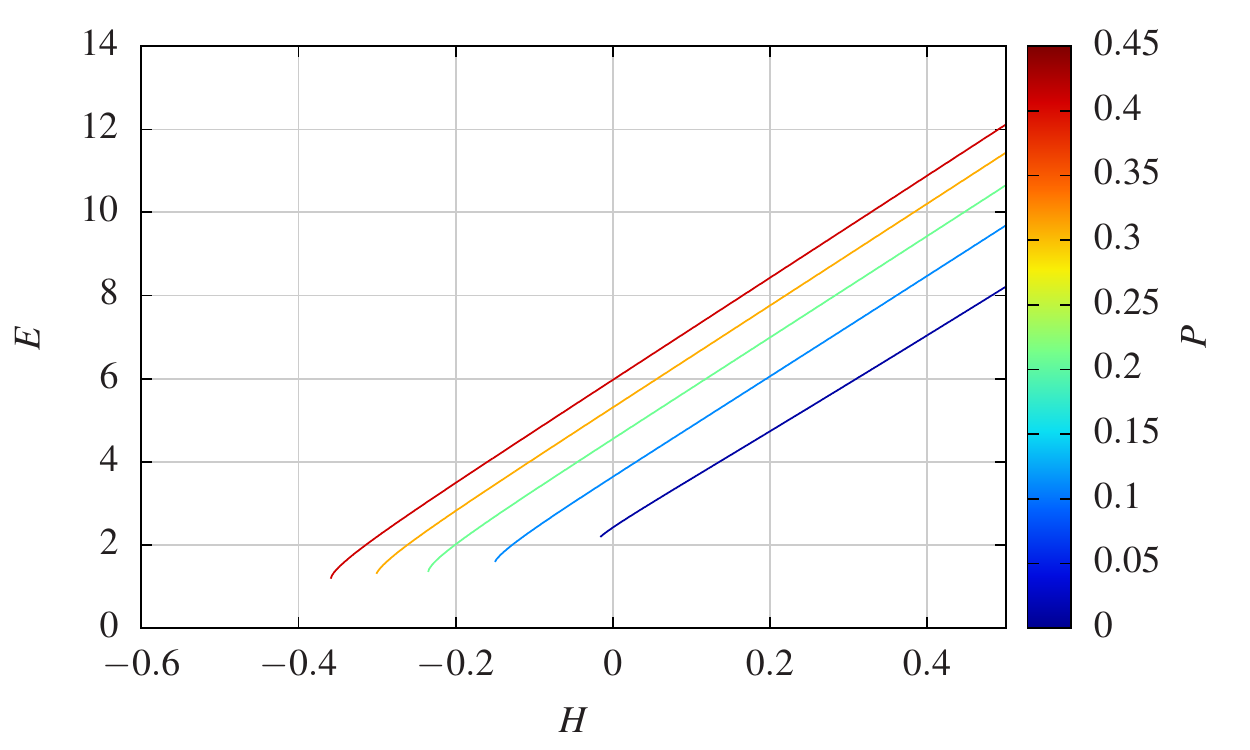}
\includegraphics[height=4.5cm]{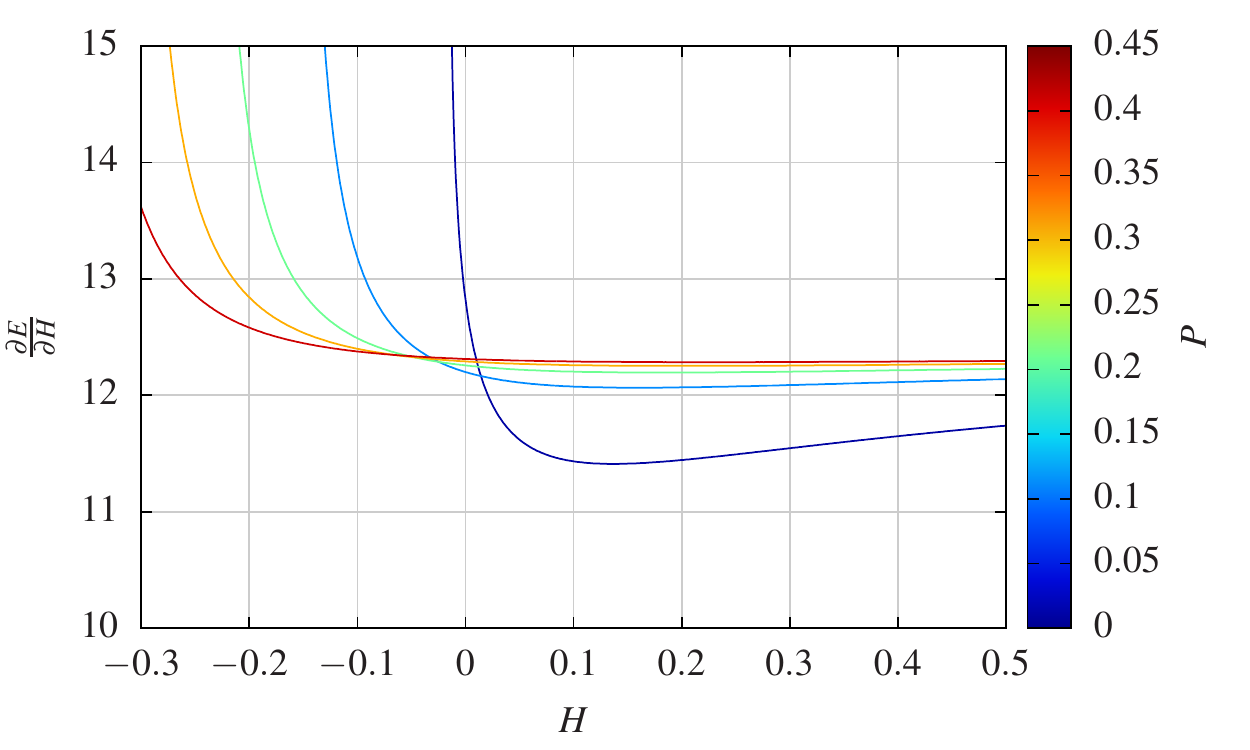}
\caption{Regularized energy and its derivative as a function of $H$ for different value of the pressure $P$. Here $g=0.2$}
\end{figure}
\be
\left. \chi_0 (P)= -\frac{\partial^2 E_{reg}}{\partial H^2} \right|_{H=0} >0.
\ee
This behavior does not change the type of the magnetic medium described by the model. It is still a ferromagnetic medium as we have a permanent magnetization, and the field wants to join the external field smoothly. It simply shows that the BPS Skyrme model with the new baby potential responds in a very nonlinear way to the external magnetic field. 
\begin{figure}
\includegraphics[height=6.5cm]{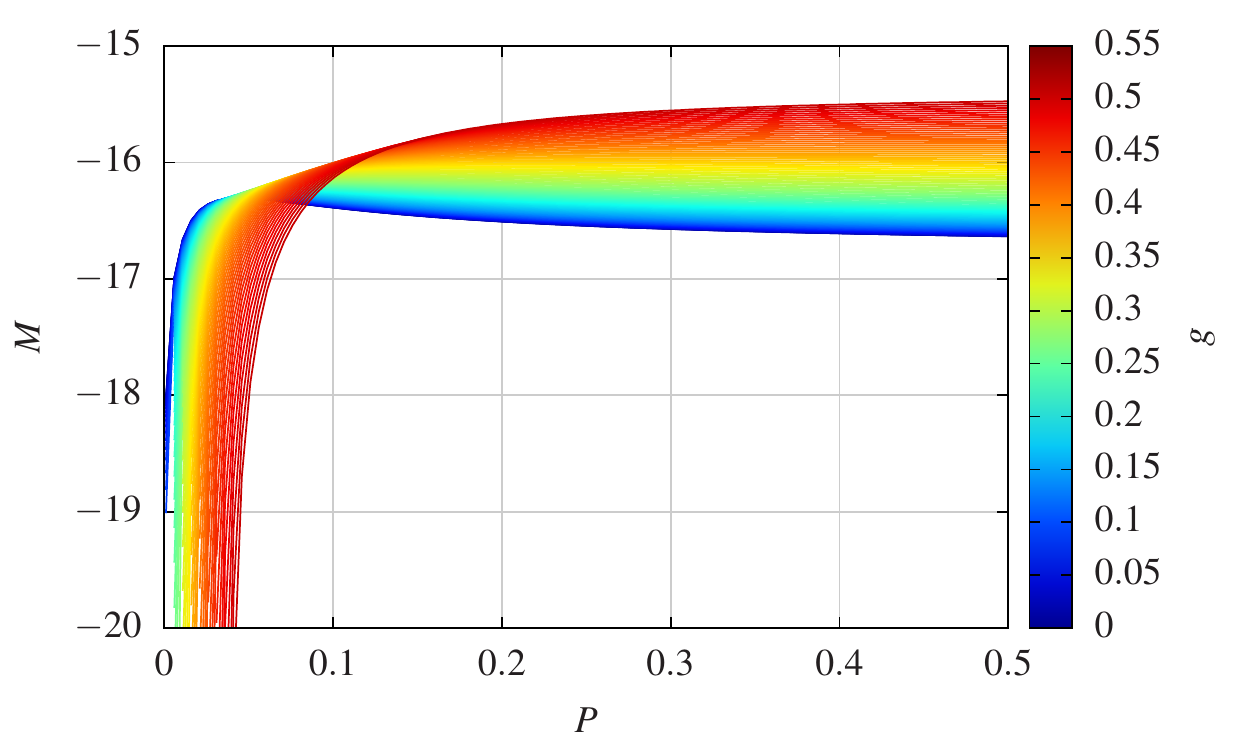}
\caption{Magnetization at $H=0$ as a function of the pressure, for different $g$}
\end{figure}
\begin{figure}
\includegraphics[height=6.5cm]{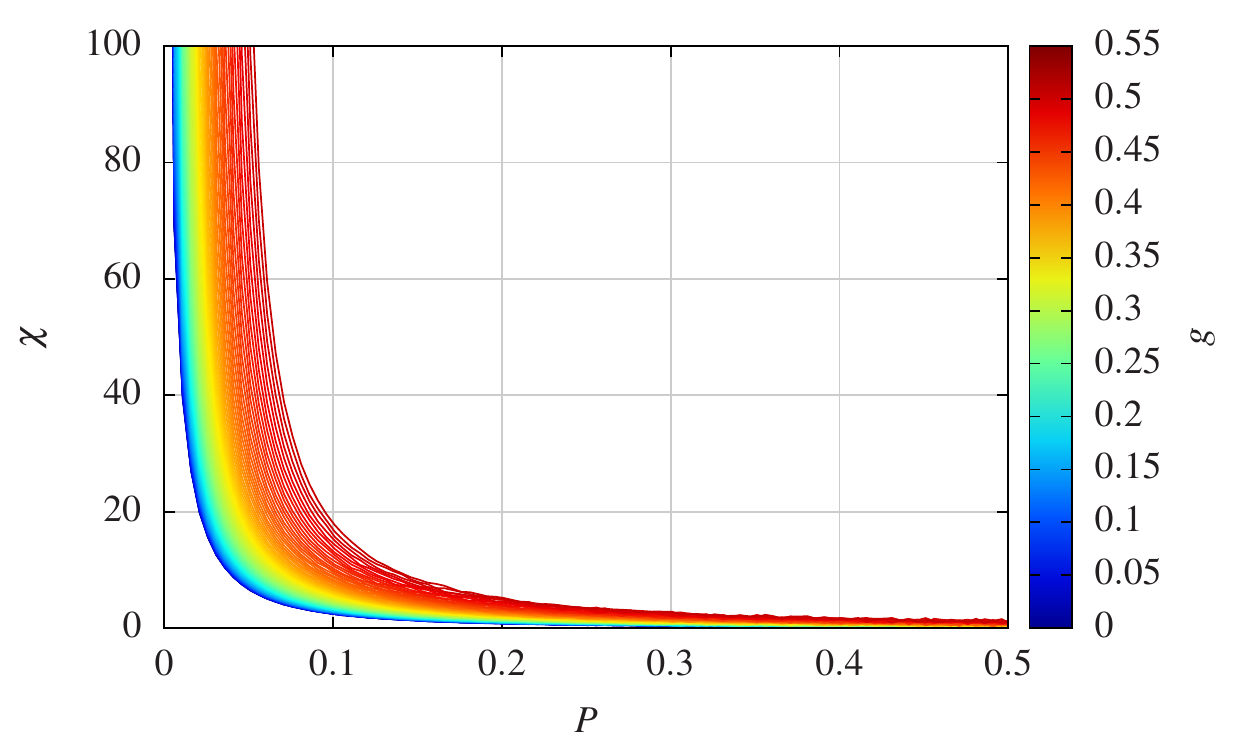}
\caption{Susceptibility at $H=0$ as a function of the pressure, for different $g$}
\end{figure}
\section{Summary}
In this paper, we have analyzed the gauged BPS baby Skyrme model with a two vacuum potential. It has been shown that although there are no baby skyrmions in the "unperturbed" case (no pressure and vanishing magnetic field at the compacton boundary), such solitons do appear if an external pressure and/or a constant external magnetic field are applied. Strictly speaking, baby skyrmions exist if these parameters are above their critical values $P_{min}, H_{min}$. The resulting critical curves (in the $P,H$ plane) have been obtained for several values of the electromagnetic coupling constant. Such a curve divides the system into two phases, a solitonic and a non-solitonic one, and crossing the curve corresponds to a phase transition from the topological to the non-topological phase. 
\\
Otherwise, the thermodynamical and magnetic properties of the model in the solitonic phase are qualitatively very similar to the previously analyzed one vacuum case. The baby skyrmions form a non-linear ferromagnetic, i.e. permanently magnetized, type of matter (except for the $P<P_{crit}(g)$ case, where the non-solitonic phase is reached as $H\rightarrow 0$). 
\begin{figure}
\includegraphics[height=6.5cm]{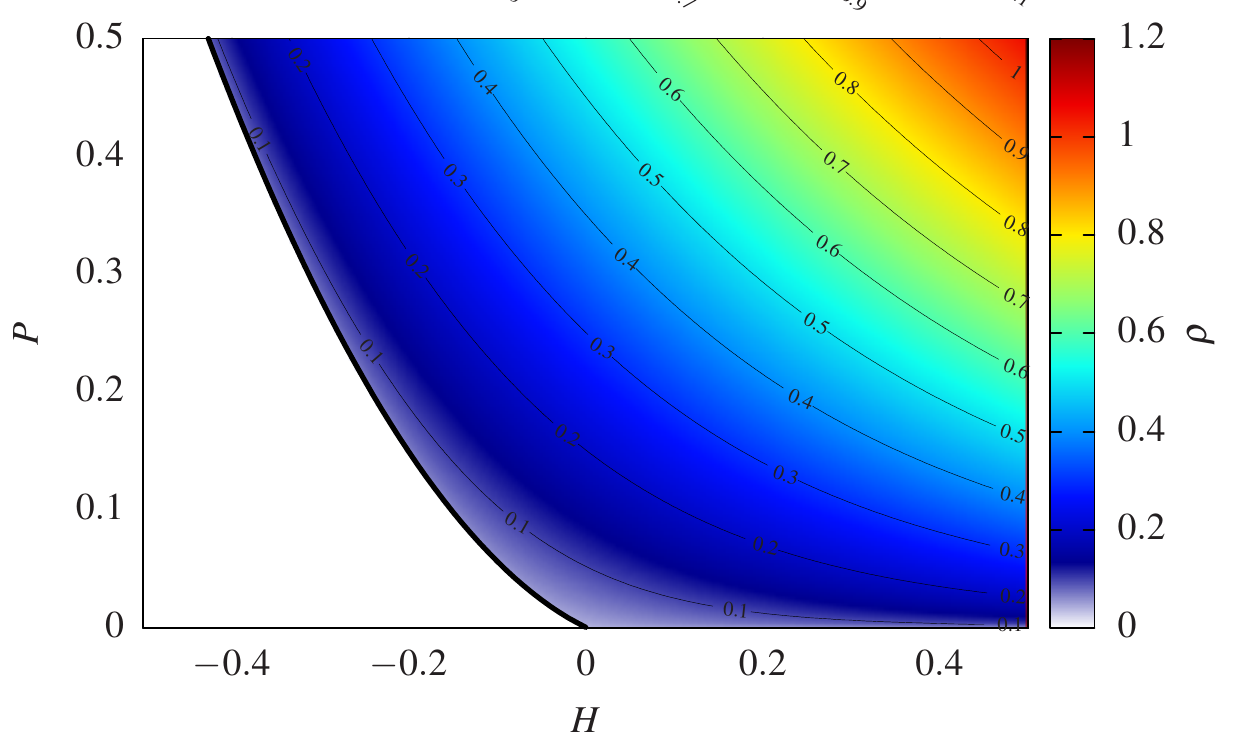}
\caption{Average density for $g=0.1$. The white region denotes the non-solitonic phase.}
\end{figure}
Let us note that a topology changing phase transition is a known phenomenon in Skyrme models. For example, a transition between phases with skyrmions and half-skyrmions was proposed in the context  of dense nuclear matter \cite{rho} (neutron stars). Specifically, the crystal phase formed by skyrmions transforms, as the density of the matter {\it increases}, into a different crystal phase where the building blocks are half-skyrmions, instead. In our case, however, the approach to the phase transition is quite different. Namely, one can start in the topological phase and then, assuming a fixed value of the external magnetic field $H$, reduce the pressure until one reaches the non-topological phase. As the pressure decreases, the size of the skyrmions grows (pressure always squeezes the solitons in the BPS Skyrme models), and the average energy density ($ \rho_{average} \equiv E_{reg}/V$) decreases, Fig. 13. Hence, the non-skyrmion phase occurs for lower densities. In other words, in the model with the new baby potential, the BPS baby skyrmions appear only if the medium has a sufficiently high density. 

There are several directions in which the present investigation may be continued. For example, one may be interested in how the magnetic and thermodynamic properties of the gauged BPS baby skyrmions are modified by the (iso)spin. The isospinning BPS baby skyrmions are known \cite{GP}, and it would be interesting to look for such solutions also in the gauged version of the model (Isospinning baby skyrmions and 3+1 dim skyrmions have been recently considered in \cite{isorot} and \cite{mar}). 

Unfortunately, there are a lot of problems and modifications which must be solved before we could extrapolate these results to the 3+1 dimensional (near) BPS Skyrme model. First of all, the planar case is a bit special with respect to the magnetic field. Indeed, here $B$ is a pseudo-scalar instead of a pseudo-vector. Moreover, in contrast to to the planar case, the gauging of the BPS Skyrme model will break the SDiff invariance in the static sector. Furthermore, the proper, QCD-induced, coupling to the abelian gauge field requires also the Wess-Zumino-Witten term, which leads to a non-zero electric field inside the skyrmions. Finally, the inclusion of non-BPS terms, i.e., the sigma model term and the Skyrme term, may modify the magnetic properties of the system. In any case, the first step should be a derivation of the energy minimiser in the non-gauge near BPS Skyrme model, which is a difficult numerical issue \cite{mm} (for some recent numerical result in the Skyrme model with the sextic term see \cite{bjarke}, \cite{bjarke2}).

Moreover, there exists the possibility that the gauged BPS baby Skyrme model is the proper effective description of some real physical 2+1 dimensional systems. Such systems should form an almost BPS liquid with an at least approximate realisation of the SDiff symmetry at low temperature. Moreover, perturbative excitations around the vacuum, governed by the relative strength of the potential part of the BPS baby model, on the one hand, and the non-BPS addition, i.e., the usual sigma model part, on the other hand, should possess rather large masses. In fact, there is a vast literature on planar solitons of the baby skyrme type topology in a magnetized medium (see, e.g., \cite{naga-toku} and the references quoted there). The relevant topological order parameter is a local magnetization which can interact with an external magnetic field. It would be very desirable to rewrite the gauged baby BPS Skyrme model in terms of magnetization and compare with such systems. This issue is under current investigation. 
\section*{Acknowledgement}
The authors acknowledge financial support from the Ministry of Education, Culture, and Sports, Spain (Grant No. FPA2011-22776), the Xunta de Galicia (Grant No. INCITE09.296.035PR and Conselleria de Educacion), the Spanish Consolider-Ingenio 2010 Programme CPAN (CSD2007-00042), and FEDER. Further, the authors acknowledge support from the Polish FOCUS grant (No. 42/F/AW/2014).  
CN thanks the Spanish Ministery of
Education, Culture and Sports for financial support (grant FPU AP2010-5772).


\begin{thebibliography}{bib}
\bibitem{baby} B.M.A.G. Piette, B.J. Schroers and W.J.
Zakrzewski, Z. Phys. C {\bf 65} (1995) 165; B.M.A.G. Piette, B.J. 
Schroers and W.J. Zakrzewski, Nucl. Phys. B {\bf 439} (1995) 205.
\bibitem{karliner} M. Karliner, I. Hen, Nonlinearity {\bf 21} (2008)
399-408;  M. Karliner, I. Hen, arXiv:0901.1489.
\bb{grav} Y. Brihaye, T. Delsate, N. Sawado, Y. Kodama, Phys.Rev. D82 (2010) 106002; T. Delsate, M. Hayasaka, N. Sawado, Phys.Rev. D86 (2012) 125009.
\bb{jay1}
J. Jaykka, M. Speight, P. Sutcliffe, Proc. Roy. Soc. Lond. A{\bf 468}, 1085
(2012).
\bb{jay2}
J. Jaykka, M. Speight,  Phys. Rev. D{\bf 82},125030 (2010).
\bb{foster1}
D. Foster, Nonlinearity {\bf 23}, 465 (2010).
\bb{foster2} D. Foster, P. Sutcliffe, Phys.Rev. {\bf D79} (2009) 125026. 
\bb{isorot} R. A. Battye, M. Haberichter, Phys. Rev. {\bf D88}, 125016 (2013); A. Halavanau, Y. Shnir, Phys. Rev. D88, 085028 (2013).
\bb{nitta} M. Kobayashi, M. Nitta, Phys.Rev. {\bf D87} (2013) 125013; M. Nitta, Phys.Rev. {\bf D87} (2013) 025013.   
\bb{stef} S. Bolognesi, P. Sutcliffe, J.Phys. A{\bf 47} (2014) 135401.
\bb{jenn} P. Jennings, P. Sutcliffe, J.Phys. {\bf A46} (2013) 465401. 
\bb{shnir} B.A. Malomed, Y. Shnir, G. Zhilin,  Phys.Rev. {\bf D89} (2014) 085021. 
\bb{sut2014} P. Salmi, P. Sutcliffe, J. Phys. A{\bf 48} (2015) 035401.
\bibitem{skyrme} T.H.R. Skyrme, Proc. Roy. Soc. Lon. {\bf 260},
127 (1961); Nucl. Phys. {\bf 31}, 556 (1961); J. Math. Phys. {\bf
12}, 1735 (1971).
\bb{nearBPS}
C. Adam, C. Naya, J. Sanchez-Guillen, A. Wereszczynski, Phys. Rev. Lett. {\bf 111} (2013) 232501; Phys. Rev. {\bf C88} (2013) 054313.
\bb{SutBPS}
P. Sutcliffe,
JHEP {\bf 1008}, 019 (2010);
P. Sutcliffe,
JHEP {\bf 1104}, 045 (2011).
\bb{BPS}
C. Adam, J. Sanchez-Guillen, A. Wereszczynski,
Phys. Lett. B{\bf 691}, 105 (2010); 
C. Adam, J. Sanchez-Guillen, A. Wereszczynski,
Phys. Rev. D{\bf 82}, 085015 (2010).
\bb{Marl} E. Bonenfant, L. Marleau, Phys.Rev. {\bf D82} (2010) 054023; E. Bonenfant, L. Harbour, L. Marleau, Phys.Rev. {\bf D85} (2012) 114045; M.-O. Beaudoin, L. Marleau, Nucl. Phys. {\bf B883} (2014) 328.
\bibitem{Sp2} J.M. Speight, arXiv:1406.0739.
\bibitem{term} C. Adam, C. Naya, J. Sanchez-Guillen, M. Speight, A. Wereszczynski, Phys. Rev. {\bf D90} (2014) 045003. 
\bibitem{star} C. Adam, C. Naya, J. Sanchez-Guillen, R. Vazquez, A. Wereszczynski, arXiv:1407.3799.
\bb{GP}
T. Gisiger, M.B. Paranjape, Phys. Rev. D{\bf 55}, 7731 (1997).
\bb{restr-bS}
C. Adam, T. Romanczukiewicz,  J. Sanchez-Guillen, A. Wereszczynski, 
Phys. Rev. D{\bf 81}, 085007 (2010).
\bb{Sp1}
J.M. Speight, 
J. Phys. A{\bf 43}, 405201 (2010). 
\bb{stepien} L. Stepien, arXiv:1204.6194; arXiv:1205.1017. 
\bb{near-baby} S. Bolognesi, W. Zakrzewski, arXiv:1407.3140.
\bb{BPS-g}  C. Adam, C. Naya, J. Sanchez-Guillen, A. Wereszczynski, Phys. Rev. {\bf D86} (2012) 045010. 
\bb{BPS-g2} 
C. Adam, T. Romanczukiewicz,  J. Sanchez-Guillen, A. Wereszczynski, JHEP {\bf 1411} (2014) 095.
\bb{susyBPS}
C. Adam, J.M. Queiruga, J. Sanchez-Guillen, A. Wereszczynski,
Phys. Rev. D{\bf 84} (2011) 025008; JHEP {\bf 1305} (2013) 108. 
\bb{nitta-susy}
M. Nitta, S. Sasaki, Phys. Rev. D{\bf 90} (2014) 105001.
\bb{wit1}
E. Witten,
Nucl. Phys. B{\bf 223}, 422 (1983);
Nucl. Phys. B{\bf 233}, 433 (1983);
C.G. Callan, E. Witten,
Nucl. Phys. B{\bf 239}, 161 (1984).
\bb{g-Skyrme1} B.M.A.G. Piette, D. H. Tchrakian, Phys. Rev. {\bf D62}, 025020 (2000).
\bb{g-Skyrme2} E. Radu, D. H. Tchrakian, Phys. Lett. {\bf B 632}, 109 (2006).
\bb{foster} D. Foster, D. Harland, work in progress. 
\bb{shnir1} Ya. Shnir, G. Zhilin,  Phys.Rev. {\bf D89} (2014) 105010.
\bb{bjarke} S. B. Gudnason, M. Nitta, Phys. Rev. {\bf D89}, 025012 (2014); S. B. Gudnason, M. Nitta, Phys.Rev. {\bf D90} (2014) 085007.
\bb{GPS}
J. Gladikowski, B.M.A.G. Piette, B.J. Schroers,
Phys. Rev. D{\bf 53} 844, (1996).
\bb{schr1}
B.J. Schroers,
Phys. Lett. B{\bf 356}  291, (1995).
\bb{rho} H.-J. Lee, B.-Y. Park, D.-P. Min, M. Rho, V. Vento, Nucl. Phys. {\bf A 723}, 427 (2003);  Y.-L. Ma, M. Harada, H. K. Lee, Y. Oh, B.-Y. Park, M. Rho, Phys. Rev. {\bf D 88}, 014016 (2013); M. Rho, S.-J. Sin, I. Zahed, Phys. Lett. {\bf B 689}, 23 (2010).
\bb{mar} R. A. Battye, M. Haberichter, S. Krusch, arXiv:1407.3264; M. Haberichter, arXiv:1411.4142.
\bb{mm} M. Gillard, M. Speight, work in progress.
\bb{bjarke2} S. B. Gudnason, M. Nitta, arXiv:1410.8407. 
\bb{naga-toku}
N. Nagaosa, Y. Tokura, Nature Nanotechnology {\bf 8} (2013) 899.

\end{thebibliography}
\end{document}